\documentclass[aip,jmp,amsmath,amssymb,reprint]{revtex4-2}

\usepackage{graphicx}
\usepackage{dcolumn}
\usepackage{bm}
\usepackage{color}

\usepackage[utf8]{inputenc}
\usepackage[T1]{fontenc}
\usepackage{mathptmx}
\usepackage{etoolbox}

\begin{document}

\preprint{AIP/123-QED}

\title
{Does (TaSe $ _4 $) $ _ 2 $I really harbor an axionic charge density wave ? }

\author{A.A. Sinchenko}
\email{aasinch@mail.ru.}
\affiliation{M.V. Lomonosov Moscow State University, 119991, Moscow, Russia}
\affiliation{Kotelnikov Institute of Radioengineering and Electronics of RAS, 125009 Moscow, Russia}%

\author{R. Ballou}
\author{J.E. Lorenzo}
\author{Th. Grenet}
\author{P. Monceau}
\email{pierre.monceau@neel.cnrs.fr}
\affiliation{Univ. Grenoble Alpes, Inst. Neel, F-38042 Grenoble, France, CNRS, Inst. Neel, F-38042 Grenoble, France}%

\date{\today}

\begin{abstract}
A recent experimental work has reported an excess of the non-linear conductivity in the charge density wave (CDW) sliding mode of the quasi one-dimensional compound (TaSe$_4$)$_2$I when a magnetic field is applied co-linearly to the electric field (J. Gooth, \textit{et al.}, Nature 575, 315 (2019). This result has opened a new conceptual approach where the CDW gap in (TaSe$_4$)$_2$I is opened between Weyl fermions of opposite chirality with the assumption that this compound is a Weyl semi-metal in its undistorted high temperature phase. We report measurements in the sliding state of (TaSe$_4$)$_2$I performed in similar conditions. We have found no increase of the magnetoconductivity. In our attempts for understanding this unsettling discrepancy we stress the specific nature of the Peierls transition in (TaSe$_4$)$_2$I and the strong electron-phonon coupling present in this compound. Given the lack of further evidences, we think that it is premature to assert that (TaSe$_4$)$_2$I is an axionic insulator.      
\end{abstract}

\maketitle

%\section{Introduction}

Experimental and theoretical investigations in condensed matter can allow the realization of quantum field theory predictions such as, for example, the axion electrodynamics\cite{Peccei77}. The axion particle has become a candidate responsible for some or all of the dark matter of the universe\cite{Preskill83}. The search for experiment detection of axion has been recently reviewed \cite{Sikivie21} and new experiment proposal is underway\cite{Grenet21}.

Since the discovery of three-dimensional (3D) topological insulators\cite{Hasan10}, a large interest has been devoted to Weyl semi-metals (see Ref.~\onlinecite{Lv21} for a recent review). They are materials in which low energy electronic quasi-particles behave as chiral relativistic fermions without rest mass known as Weyl fermions. Weyl fermions exist at isolated crossing points between conduction and valence bands, called Weyl nodes. Axion electrodynamics in Weyl semi metals arise from the separation of Weyl nodes of opposite chirality in momentum space and energy. Turning on interactions, chiral symmetry breaking is the phenomenon of the spontaneous generation of an effective mass of Weyl fermions, namely a pairing between the fermions (electrons) and holes with different chiralities which opens gaps at the Weyl points and induces a charge density wave (CDW). The opening of this gap transforms the Weyl semi-metal into an axion insulator and the phase of the CDW is identified to the dynamic axion field $\theta$ \cite{Wang13}. The CDW phason is the axion mode and its dynamics is therefore described by the topological term $\theta E \cdot B$. The CDW phason mode is also known as the Goldstone mode. Applying $B$ collinearly to $E$ leads to an extra chiral flow of charges  which can be detected from magnetoelectric measurements. 

Recently a  gigantic negative longitudinal magnetoresistance (LMR) was observed \cite{Gooth19} in the CDW compound (TaSe$_4$)$_2$I when the magnetic field co-linear to the electric field is applied parallel to the CDW direction. Such LMR appears only in the state of collective CDW motion (sliding) at electric fields larger than the characteristic threshold electric field, $E_t$. The authors concluded that the LMR results from a chiral anomaly %, which is most of remarkable effect arising in Weyl semi-metals, 
appearing in the Peierls state of (TaSe$_4$)$_2$I and originating from the anomalous axionic contribution %of the chiral anomaly 
to the phason current. % in the configuration of the parallel alignment of the electric and magnetic fields. 
This result was considered as the first determination of the occurrence of axion electrodynamics in condensed matter \cite{Wang13,Sekine21}, reinforcing the new concept of axionic charge density wave \cite{Nenno20}.

%However any result, especially when it is unusual, needs to be confirmed. 
%In a yet single independent work \cite{ZZ20} an attempt was made to find an experimental prove of LMR effects in (TaSe$_4$)$_2$I. 
Oddly, and in contrast to the results reported in Ref. \onlinecite{Gooth19}, the LMR measured by Cohn \textit{et al.} \cite{ZZ20} was positive and does not exceeded a fraction of per cent in the sliding as well as in the static regimes of the CDW in the temperature range down to 30K. The main difference between both experiments is the sample geometry and the position of the contacts: in Ref. \onlinecite{Gooth19} massive single crystals with cross-section of order $10^3-10^4$ $\mu$m$^2$ and the length between the potential contacts 3-10 mm were used. While in Ref. \onlinecite{ZZ20} very thin samples in the order 10 $\mu$m$^2$ of cross-section and distances between the edges of the contacts of order 10$^2$ $\mu$m. 

In view of this strong disagreement we have repeated the measurements of the non-linear state of (TaSe$_4$)$_2$I under magnetic field. To that end we have set experimental conditions  mostly identical to those in Ref. \onlinecite{Gooth19} using nearly the same sample geometry, the same contact preparation method and the same measurement setup. We have not found any significant contribution of the magnetic field in the sliding motion of the CDW. We discuss the possible reasons of the discrepancy including the generation of CDW dislocations under a strong thermal gradient and we stress the role of the strong electron-phonon coupling in (TaSe$_4$)$_2$I.

%\section{Description of the compound}

(TaSe$_4$)$_2$I is a chain-like material displaying a Peierls phase transition (for a review on the Peierls transition and compounds, see Refs. \onlinecite{Gruner,Monceau12}) at T$_P\approx$263K, this value varies rather strongly, that has been intensively studied since several decades \cite{Gressier82,Gressier84}. (TaSe$_4$)$_2$I crystallizes with tetragonal symmetry (space group I$422$) and consists of TaSe$_4$ Van der Waals chains parallel to $c$-axis separated by iodine atoms. In an TaSe$_4$ infinite chain, each Ta is sandwiched by two nearly rectangular selenium units forming a Se$_2^{-2}$ pair. The dihedral angle between adjacent rectangles is almost 45$^\circ$ (44$^\circ$ and 46$^\circ$), so the Ta atom is located at the center of a rectangular anti-prism of eight Se atoms\cite{Gressier84}. The interaction between Ta atoms is through $d_{z^2}$ overlap along the chain (the shortest interchain metal–metal distance is about 6.7\AA{} to be compared to the intrachain average value $d$ of about 3.2\AA{}.

Band structure calculations \cite{Gressier84A} suggest a single $d_{z^2}$ electronic band at the Fermi level. This band is 1/4 filled with one free electron per Ta$^{+4}$ Ta$^{5+}$ 4Se$^{2-}$ 2I$^-$ formula unit. Consecutive Ta atoms occupy two alternating non-equivalent sites, but the Ta-Ta distance is unique, d$_{Ta-Ta}=$3.206\AA{}. Owing to the Se$_4$-unit rotation pattern, the crystallographic unit cell parameter is $c=4d_{Ta-Ta}$. On that basis, one may expect (TaSe$_4$)$_2$ to be an insulator, because the Fermi wave vector corresponds to a Brillouin zone boundary:                          $k_F=1/4(\pi/d_{Ta-Ta})=\pi/c=c^*/2$. Assuming that Coulomb repulsion favors an antiphase arrangement of CDWs on 
adjacent chains, one expects the CDW satellites below $T_P$ to appear at or near $c^*+$ and $-a^*$ or $c^*+$ and $-b^*$ close to the allowed Bragg reflections, $G$, of the parent undistorted structure. Indeed electron and X-ray diffraction experiments\cite{Lee87} below $T_P$ show the appearance of eight first-order satellites around each room-temperature reflection at incommensurate positions of the type $G+$ ($\pm\delta h$, $\pm\delta k$, $\pm\delta l$) close to each main Bragg reflection $G=(H, K, L)$ with $\delta h=\delta k=0.045$ and $\delta l=0.085$, effectively very near of the Brillouin zone center. These eight satellites correspond to four degenerate domains with a single incommensurate modulation, single $q$-domains with orthorhombic and identically stable monoclinic symmetry \cite{Zhang20}. 

The atomic displacements of the modulated CDW state consist of two parts: the major one \cite{Lee87,Lorenzo98} is a transverse acoustic-like wave (perpendicular to $q$) with equal amplitudes of all atoms (0.13\AA{}) and a new periodicity involving mostly $z$-polarized Ta displacements that corresponds to a LLSS pattern of long and short in-chain Ta-Ta distances (Ta-tetramerization modes) which represents the CDW. The amplitude of the displacements of this tetramerization is one order of magnitude smaller than that of the acoustic component, and was only detectable by using X-ray anomalous scattering techniques \cite{Favre01}. Since the electronic variables are unlikely to couple directly to such a long-wavelength acoustic shear wave, a model has been proposed \cite{Lorenzo98} in which the soft Ta-tetramerization modes interact with the acoustic degrees of freedom and induce the condensation of a mixed acoustic/optic ionic modulation. In this model, the values of the satellite wave vector components are not related to the topology of the Fermi surface (FS). They are determined by the strength of the gradient interaction terms between optical and acoustic degrees of freedom. Similarly to models developed in the context of incommensurate long-wavelength modulated dielectrics such as quartz, the incommensurate structure arises from the presence of a pseudo-Lifshitz invariant involving an optical order parameter and the elastic deformations \cite{Lorenzo98}. The presence and further detection of the Ta-tetramerization \cite{Favre01} provides the key for understanding the mechanism of the phase transition at $T_P$ which can be described as a Brillouin zone center Peierls instability.

The FS of (TaSe$_4$)$_2$ has been mapped by angle-resolved photoelectron spectroscopy (ARPES) \cite{Hufner99}. %IFS mapping based on measurements of the angular photoelectron intensity distribution (ARPES) has yielded \cite{Hufner99} the determination of the FS of (TaSe$_4$)$_2$I. 
It consists of parallel planes oriented perpendicularly to the chain direction. The dispersion of the $d_{z^2}$ band has also been recorded \cite{Voit00} at $T=300$ K for a range of wavevectors along the 1D chain direction. The band shows a strong dispersion throughout the first Brillouin zone with minimum at $\Gamma$, the centre of the Brillouin zone, also detectable with a weaker intensity in the second and third Brillouin zones. The low intensity at $E_F$ for wave vectors close to $k_F$ is the indication of a pseudogap still present above the Peierls transition temperature. The temperature dependence of the optical conductivity exhibits a strong absorption due to the CDW gap. That gap is already present in the electronic excitation spectrum above the Peierls transition manifesting that, even at high temperature free carriers are condensed into a fluctuating CDW ground state. However, above $T_P$, as shown by neutron\cite{Lorenzo98} or X-rays scattering \cite{Lee87}, the CDW does not exhibit long-range order, but has the physical properties of a liquid. Below $T_P$ this “liquid” condensate crystallizes into a superlattice structure. Optical spectra show a nearly continuous change between 15 and 300K. The shape of the longitudinal optical conductivity spectrum exhibits just a continuous narrowing through the Peierls transition. Recent ARPES measurements are also in agreement with the gapless nature of (TaSe$_4$)$_2$I at room temperature and reveal its characteristic dispersion \cite{Li21}. 

%\section{Experiments}

Single crystals of (TaSe$_4$)$_2$I were grown by direct combination of the elements in sealed quartz tubes in a gradient furnace at a temperature ranging from 850-900 C with iodine as transport agent \cite{Gressier82}. For the experiments we selected single crystals with nearly the same size as those studied in Ref. \onlinecite{Gooth19}. Contacts for electrical transport measurements in the four-probe configuration have been prepared using thin gold wires glued on the sample by silver paste. All current-voltage characteristics (IVc) and resistance measurements were carried out in the current mode. To exclude possible heating effects a part of the sample was covered by highly heat-conducting epoxy as it was done in Ref. \onlinecite{Gooth19} and another part of sample was covered by apiezon grease as we usually do it for such experiments. Magnetotransport measurements were performed at two mutually perpendicular orientations of the magnetic field in the field range up to 8 T using a superconducting solenoid. 

\begin{figure}
	\includegraphics[width=8.5cm]{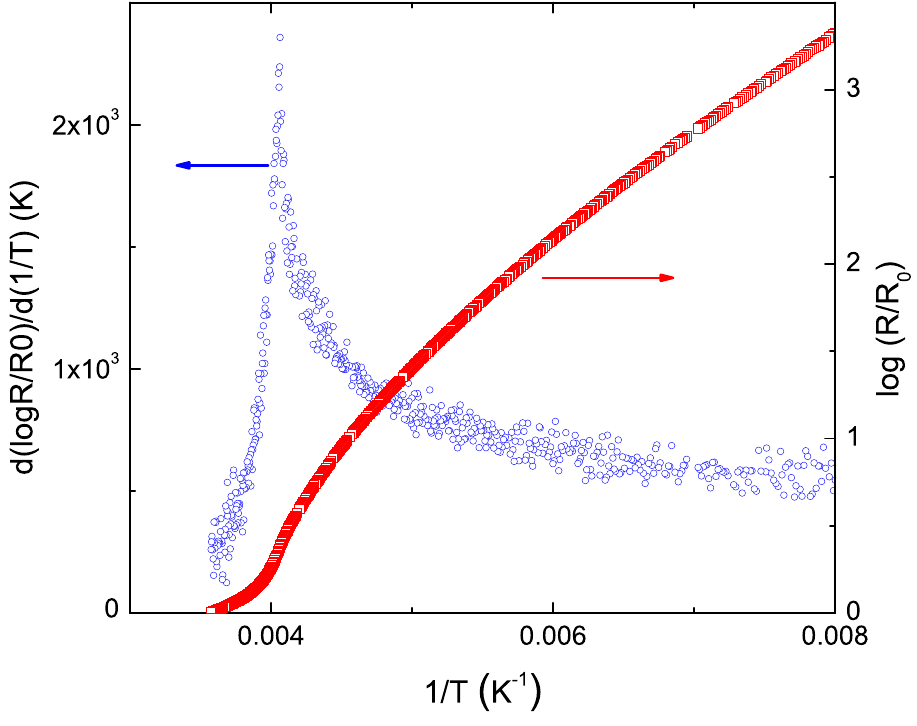}
	\caption{\label{F1} Logarithm of the electrical resistance, $R$, normalized to the $R_0 = R$ (300 K) (right axis, red) and its logarithmic derivative (left axis, blue) as a function of the inverse temperature.}
\end{figure} 

The temperature dependence of the three studied samples displayed the same behavior: a strong monotonic grow of the resistance with decrease of temperature (Fig. \ref{F1} right axis). A pronounced peak is observed in the logarithmic derivative, $d[log(R/R_0)]/d(1/T)$ versus the inverse temperature (Fig. \ref{F1} left axis). The position of this peak determines the transition into the CDW state \cite{Monceau85,Wang83} the value of which is $T_P=248$ K in our samples. It has been often argued that differences in $T_P$ values reflect small deviations from stoechiometry (loss of iodine). In that case one expects the change in the value of $T_P$ to be correlated with a change in the conduction electron Fermi wavevector and consequently in the satellite position. No such correlation was observed \cite{Lorenzo98}. We shall come back to this point below.

\begin{figure}
	\includegraphics[width=8.5cm]{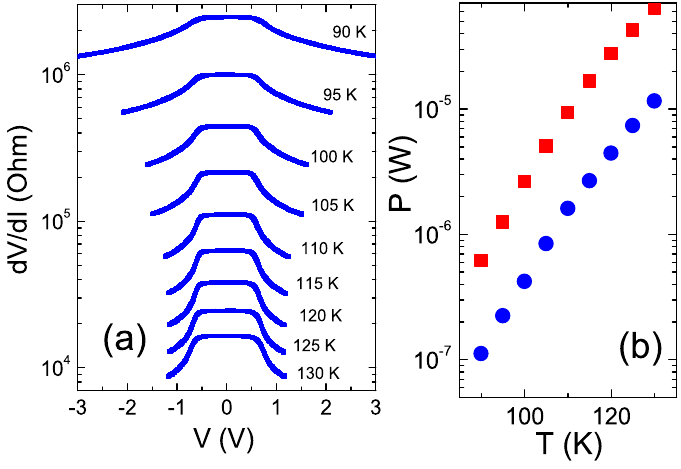}
	\caption{\label{F2} (a) Differential current-voltage characteristics, $dV/dI(V)$, in one of the samples demonstrating the CDW collective motion in the temperature range 90-130 K; (b) Joule power dissipated at the threshold electric field, $P(E_t)$, (blue circles) and at twice the threshold, $P(2E_t)$, (red squares) as a function of temperature.}
\end{figure}

Measurements of IVc on the samples covered by epoxy demonstrated a strong monotonic decrease of the resistance with increase of the current without exhibiting a strong threshold behavior at the onset of the CDW sliding regime, probably due to Joule heating. We only observed sharp threshold characteristics for samples covered by apiezon. Fig. \ref{F2} (a)  shows a set of such characteristics for one of the samples in the temperature range 90-130 K. In Fig. \ref{F2} (b) we show the temperature dependence of the Joule power, $P$, dissipated at the threshold voltage (blue circles) and at twice the threshold voltage (red squares). It can be seen that Joule power increases as the temperature increases, qualitatively in the same way as it was reported in Ref. \onlinecite{Gooth19} but with a difference of near three orders of magnitude.

\begin{figure}
	\includegraphics[width=8.5cm]{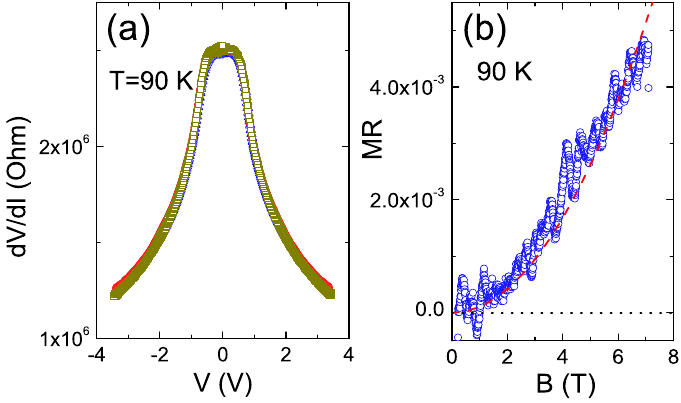}
	\caption{\label{F3} (a) Differential current-voltage characteristics, $dV/dI(V)$, at $T=90$ K in magnetic field $B=0$ T (dark yellow symbols), magnetic field $B=7$ T parallel to the current (red symbols) and $B=7$ T perpendicular to the current (blue symbols); (b) Magnetoresistance of the same (TaSe$_4$)$_2$I sample at $T=90$ K in a magnetic field parallel to the current $I=1.2I_t$ where $I_t$ is the threshold current for the CDW sliding.}
\end{figure}

The main experimental result of the present work is obtained from measurements under magnetic field, $B$. Fig. \ref{F3} (a) shows differential IVc at T = 90 K at $B=0$ T (green), at $B=7$ T (red) with the magnetic field aligned parallel to the $c$-axis (chain direction) and collinear with the electric field, $E$, and in the transverse orientation (blue) with $B$ and $E$ perpendicular. It can be seen that the effect of magnetic field is negligibly small as it was also found in Ref. \onlinecite{ZZ20}. Moreover, the magnetoresistance is positive both for transverse and longitudinal orientation of $B$. As an example Fig. \ref{F3} (b) shows the result of the direct measurement of magnetoresistance, $MR=\dfrac{R(B)-R(0)}{R(0)}$, swiping B from 0 to 7T in the case  of $B$ parallel to $I$ with the applied current of $I=6\times10^{-4}$ mA just corresponding to the sliding state of the CDW. It can be seen that MR is positive, very small (of order $10^{-3}$) and follows a quadratic dependence on magnetic field.

%\section{Discussion}

Gooth \textit{et al.} \cite{Gooth19} have measured a longitudinal MR in the sliding state of (TaSe$_4$)$_2$I which has been considered as the first experimental observation of the axion dynamics in condensed matter. Our results, as well as those of Ref.~\onlinecite{ZZ20}, do not show any manifestation of magnetoresistance. % and therefore contradicts this finding. %What can be the reasons of such a contradictory result ? In the following we open some lines: the nature of the symmetry breaking which opens the gap in the excitations and yield the formation of a CDW, the predominant role of the e-ph interaction which has been ignored or at least under-estimated, the specific temperature dependence of the excess of conductivity in the sliding state and its correlation with heating effects, the specific conditions for a possible sliding due to axion electrodynamics and the CDW defects in the sliding mode…
In the following we discuss some points that may serve to illustrate the difficulty to reconcile the actual features of (TaSe$_4$)$_2$I with the Weyl semi-metal physics and the axion electrodynamics.

%\subsection{The CDW modulation}

%In the more extensive paper, 
(TaSe$_4$)$_2$I was characterized as a semi-metal which undergoes a phase transition by breaking of the chiral symmetry between Weyl nodes (WP). From first principles calculations (TaSe$_4$)$_2$I is described as a semi-metal \cite{Shi21} with coupled WP by e-e interactions that yield the opening of a gap and the onset of a CDW. It is found that the entire Fermi surface is formed from topological bands connecting bulk chiral fermions. In the electronic structure 48 WPs within 15 meV of $E_F$ are found. 

The same authors performed X-ray diffraction experiments in their samples\cite{Shi21}. A set of 8 satellites, appearing at positions ($\pm \eta, \pm \eta, \pm \delta$), are obtained at $\eta=0.027$, and $\delta=0.012$ and $T_P=248$ K. However this CDW modulation wavevector is much shorter than the nesting vectors between WPs. These nesting vectors can be approximated to the closest integer number of the experimental CDW wave vector $h,h,k$ such as: $q_{WP}=mh,nk,ol$ with values of $m$, $n$ and $o$ ranging from 0-20 %such as: (15,-17,6) or (-15,15,7), ….
that yield the authors \cite{Shi21} “to suggest that FWs nesting is not itself the origin of the CDW”. Thus the suggested electronic CDW does not nest the Weyl points and therefore does not break the chiral symmetry. The broken symmetry originates from e-ph coupling between pieces of the Fermi surface separated by the $q$ vector. To end this paragraph on the components of the CDW modulation wavevector, these values contrast with those reported in Ref.~\onlinecite{Lee87,Lorenzo98,Favre01}: $\eta=0.045\pm0.005$, and $\delta=0.085\pm0.005$ in the nominally pure compound. A smaller value agrees with the results in Nb-doped samples \cite{Lorenzo98}.

More importantly, the authors describe (TaSe$_4$)$_2$I as an assembly of nearly independent infinite (TaSe$_4$) chains with a very weak interchain coupling. That is not corroborated by elastic neutron scattering \cite{Lorenzo98} which show that above $T_P$ the inverse correlation lengths along the three directions $c^*$, $a^*+b^*$ and $a^*-b^*$ are nearly isotropic with an anisotropy parameter amounting to 1.8. 40K above $T_P$ the average number of correlated chains was estimated at around 200. The strong interchain interactions of the CDW, which are at variance with those typically found for other quasi-1D conductors, is also manifested in the long wavelength acoustic character of the atomic displacements, as has been explained above. This strong interaction with the underlying lattice was lacking in this work \cite{Shi21}, as well.

%\subsection{The CDW sliding state}

The possibility that the CDW can slide was suggested by Fr\"{o}hlich \cite{Frolich54} and studied by Allender et al. \cite{Allender74} in a tight binding model. If the lattice wave moves with the electrons with velocity $v_s$  the CDW order parameter will vary as: $\Delta=\mid\Delta\mid exp(2ik_F(x-v_st)$. In a 1D model the two planes of the FS are at $(-\pi/k_F)+q$ and $(+\pi/k_F)+q$ with $m^*v_s=\hbar q$ where $m^*$ is the effective electronic mass, that yield in the static reference frame a current $J=nev_s$ where $n$ is the number of electrons in the band affected by the CDW. The CDW velocity can be measured from the periodic signal superposed to the static nonlinear current, the so-called narrow band noise (NBN) or from the Shapiro steps structure when the motion of the CDW is locked by an applied external rf field. Thus the energy gap reduces the elastic scattering of individual electrons because there in no state for energy relaxation. The properties of this nonlinear state have been studied in detail for quasi-1D materials with different crystallographic symmetries where chirality does not play any role. Note that the $\hbar k_F$ term makes the difference in energies between the left- and right-band sides of the displaced distribution.

In the case in which the ground CDW state is an axion insulator, in addition to the Fr\"{o}hlich Hamiltonian, a supplementary axion term (similarly to quantum field theory) was added \cite{Gooth19,Schmeltzer21} containing the contribution $\theta E \cdot B$ of the chiral anomaly in the undistorted Weyl semi-metal, with $\theta$ being the dynamical axion field identified to the dynamics of the phase of the CDW. There is thus an extra contribution to the phason current from this chiral anomaly and an excess conductivity in the sliding state. In addition of the electric force which, above the depinning electric field from impurities, makes the CDW to slide, there is an extra force which increases the CDW velocity and therefore the nonlinear current. This effect is similar to the case where the velocity of a CDW in the sliding state is increased by the application of an independent current on a small part of the sample\cite{Saint-Lager89}. This result also illustrates the very long range coherence of the CDW motion.

%\subsection{The experimental results}

Let us consider now the experimental results of Ref. \onlinecite{Gooth19}. The non-linearity in the IV characteristics have been essentially studied in one crystal (A). The data do not exhibit a real sharp threshold for the onset of the CDW sliding, but rather a continuous decrease of resistance. More surprisingly, this non-linearity in the IV characteristics is only seen at $T<180$ K much below the Peierls transition temperature when previous results show that the sliding occurs immediately below $T_P$ with a temperature dependent $E_t$. When a magnetic field is applied collinearly with the electric field a sharper decrease of the resistivity in the non-linear state is observed at a very delimited conditions: at a single temperature, namely $T=80$ K and under a very large applied voltage. This effect is quantified by measuring the magnetoconductivity as: $\Delta(dI/dV)_B=(dI/dV)_B-(dI/dV)_{0T}$. The most disturbing result is the temperature dependence of $\Delta(dI/dV)_B$  in Fig. 3h in Ref. \onlinecite{Gooth19}. It is null down to 120K and it sharply increases at 80K. The data for samples A and B at $T=105$ K with $B$ and $E$ collinear shown in the Extended Data Fig. 8 (a), (b) for sample A and (d), (e) for sample B are very similar to our results shown in Fig. \ref{F3}. Something else occurs between 105 K and 80 K which is not explained in Ref. \onlinecite{Gooth19}.

In our work (see Fig. 2), as well as in previously published studies \cite{Monceau85,Wang83,Fleming86} in (TaSe$_4$)$_2$I, the relative resistance drop $R(E<E_t)/R(4E_t)$ at $T=80-120$ K is small, not more than a factor of 2. In contrast, in Ref. \onlinecite{Gooth19} this relation is more than 10 times larger at zero magnetic field and near 100 times larger at $B=9$ T. Gigantic heating effect in Ref. \onlinecite{Gooth19} which takes place at high bias ought to be considered. Indeed, the authors have estimated Joule heating only at threshold voltage determined as $V_t=6$ V at 80 K and obtained for $P\approx20$ $\mu$W that is really small enough. However, at $V=27$ V (near $4E_t$) this power is already $P\approx60$ mW that is more than 3 order of magnitude larger. However at 105 K these values are in the range of 200 mW. By comparison the power dissipated in our work at $T=90$ K is $P=0.1$ $\mu$W at the threshold and $\approx 3$ $\mu$W at four times the threshold. This is two order of magnitude smaller than in Ref.~\onlinecite{Gooth19} at the same condition and $T=80$ K. As a result, one may consider that the main contribution to the resistance drop in Ref. \onlinecite{Gooth19} may originate as a consequence of heating (as we discuss below) while the contribution of the CDW sliding to the total drop of resistance may be less than 10 \%. This is simply  one order of magnitude smaller than the magnetoresistance effect and reported as an indication of axion transport in (TaSe$_4$)$_2$I.

%\subsection{CDW sliding mode in a Weyl-CDW compound}

Transport properties in quasi 1D metals have been studied under the assumption that the single particle gap originates from the nesting of a single band at +$k_F$ and -$k_F$. This mechanism, the Peierls instability, stands for the large part of the CDW compounds. However a large difference occurs when the CDW order opens a gap between Weyl fermions of opposite chirality. Unlike the single-band CDW ground state, the involvement of multibands in Weyl-CDWs compounds results in a gapless CDW collective mode that does not contribute to the conductivity unless the Weyl cones are tilted \cite{Mckay21}. Considering the simplest two-band Weyl model in the undistorted phase, it consists of two small Fermi pockets around each Weyl node, separated by the wave vector $q$ (that determined by x-ray scattering to be consistent with previous results). At the CDW phase transition there is condensation of particle-hole pairs consisting of a hole in one Fermi pocket and an electron separated by momentum $q$. These electron and hole states exhibit the same velocity, unlike for a 1D CDW. %Following the description we have given above with a single band, in the Weyl model with two bands, in the momentum space one band is shifted with $+q$ and the other with minus $-q$ giving a contribution $+J$ and $-J$ which annihilate.
Thus, for the sliding mode to contribute to the DC conductivity in a Weyl-CDW, it must have an asymmetry between the velocities in different Weyl pockets. This can be accomplished by shifting the nodes in energy or inducing a tilt to the Weyl cones. Is it the case for (TaSe$_4$)$_2$I ? We note however that the sliding properties of (TaSe$_4$)$_2$I are very similar to those of other materials exhibiting CDW sliding.  

%\subsection{Power dissipated}

An important point from the experimental point of view is how Joule power dissipated in the sample can lead to a strong anisotropic magnetoresistance effect. Strong heating creates strong thermal inhomogeneities because the thermal exchange can be different at different parts of the sample. Logically, one assume that such exchange will be better from the bottom surface of the sample which is in contact with the high thermal conductivity substrate compared with the top surface of the sample. Such difference leads to a strong thermal gradient from top to the bottom of the crystal which creates a corresponding thermoelectric current. Being oriented transverse to the applied electric field this current does not give a contribution to the total drop of voltage. It will be no effect under the application of a magnetic field parallel to this thermal current ($B\perp I$ configuration) because the Lorentz force is zero for thermo-electrons. Another picture should be observed when $B\parallel I$. In this case the thermoelectric current which continuously increases with the increase of transport current is affected by the Lorentz force that leads to an additional scattering and additional heating correspondingly. 

%\subsection{CDW dislocations}

As said before, the CDW ground state has a very long range order. However the CDW cannot solely be assumed to behave elastically when the CDW velocity passes through strong and isolated sharp discontinuities. The conflict between different winding rates at the interface separating regions with different velocities is released by phase slippage. In a path surrounding these vortices the phase changes by $2\pi$. The topological defects of ICDW are dislocations very similar to vortices in superfluids and their properties are close of those of dislocation lines in crystals. There are simple cases of dislocations: screw dislocations parallel to the chains and edge dislocations perpendicular to chains  with a dislocation core with dimensions $\xi_p\xi_p$  for an edge or $\xi_p\xi_\parallel$ for a screw where $\xi_p$ and $\xi_\parallel$ are the BCS amplitude coherence lengths. The CDW can be pinned at the surface, near strong impurities in the volume, at the electrodes. It was also shown that the coherence of the CDW motion is broken when a thermal gradient is applied \cite{Ong84,Verma84,Zettl85} between electrodes with the detection of two different velocities. As the CDW order parameter is suppressed in the dislocation core, the properties of the Weyl semi-metal, which are characterized by negative MR, could partially be restored. However, in this scenario the amplitude of negative MR should be very small because this effect takes place in a relatively small part of sample. However if the properties of Weyl semi-metal are to be truly restored it should also be accompanied by a significant (at least tens of per cent) positive transverse MR that was not observed in Ref. \onlinecite{Gooth19}.

In the context of the axion insulator Wang and Zhang \cite{Wang13} have identified the CDW dislocations as axion strings. An axion string is a one-dimensional dislocation of the axion field around which the axion field $\theta$ changes by $2\pi$. Such chiral modes would carry a dissipationless current. It appears from Ref. \onlinecite{Gooth19} that the magnetoconductivity is only observed when the Joule power dissipated surpasses some threshold:  60mW at $T=80$ K, 200 mW at $T=105$ K and no effect at higher temperature. Then one may think that at this so huge power dissipated, in this system out of equilibrium, amplitude modes are excited and “hot” filaments in which the CDW order parameter is suppressed can cross at high velocity from one electrode to the other. Interaction between dislocations in the lattice and dynamical axion strings have been also studied \cite{You16}.

%\section{Conclusion}

In conclusion, the result of magneto-conductivity in the sliding state of (TaSe$_4$)$_2$I has risen a very large interest since it provides the experimental proof, which lacked, of the axion dynamics in condensed matter \cite{Gooth19}. Here we have raised some points which questions this statement. The symmetry breaking in (TaSe$_4$)$_2$I results from electron-phonon interactions and not from electron-electron ones. The distorted phase of (TaSe$_4$)$_2$I consists of the condensation of a mixed acoustic/optic ionic modulation from the interaction of a Ta-tetramerization with a transverse acoustic wave. The Peierls transition, probably of the first order, is rather an order-disorder transition than a displacive one. The new approach on (TaSe$_4$)$_2$I originates from the chirality of the I$422$ space group in which it crystallizes. The 24 Weyl points derived from first principle calculations are connected with $q$ vectors that are irrelevant with the CDW $q$-vector determined by x-rays three decades ago. Our results do not show any extra conductivity in the sliding state of (TaSe$_4$)$_2$I when magnetic and electric fields are collinear. A detail analysis of the Ref. \onlinecite{Gooth19} magneto-conductivity data shows that it has only been detected abruptly below 100 K and specifically at a single temperature, $T=80$ K (while the phase transition is at $T=263$ K). And under a very large voltage which induces a large Joule power dissipation. As a possible explanation of the discrepancy between our results as well as those from Ref. \onlinecite{ZZ20} we have suggested possible hot filaments with cores having the properties of the normal state.

We do not disregard all the new fascinating theoretical and experimental discoveries around these Weyl semimetals. We however think that, on the basis of present magneto-conductivity results, the assertion that (TaSe$_4$)$_2$I would be an axionic charge density wave is not justified and this conclusion should be considered as, at least, premature. New transport experiments are necessary for the temperature dependence of the axionic magnetoresistance in a large temperature scale and under different thermal gradients. Measurements of Shapiro steps in the sliding state will also allow to determine if the application of a collinear magnetic field increases the CDW velocity and therefore if the axion field acts on the CDW long range condensate.

The data that support the findings of this study are available from the corresponding authors upon reasonable request.

\end{document}